\documentclass[prb,twocolumn,showpacs,preprintnumbers,amsmath,amssymb,floatfix,superscriptaddress]{revtex4}
\usepackage{graphicx}
\usepackage{bm}

\begin{document}

\title
{Transient tunneling effects of resonance doublets 
in triple barrier systems}

\author{Roberto Romo}
\email[Electronic mail: ]{romo@uabc.mx}
\affiliation{Facultad de Ciencias,
Universidad Aut\'onoma de Baja California\\
Apartado Postal 1880, 22800 Ensenada, Baja California, M\'exico}

\author{Jorge Villavicencio}
\email[Electronic mail:]{villavics@uabc.mx}
\affiliation{Facultad de Ciencias,
Universidad Aut\'onoma de Baja California\\
Apartado Postal 1880, 22800 Ensenada, Baja California, M\'exico}
\affiliation{Instituto de F\'{\i}sica,
Universidad Nacional Aut\'onoma de M\'exico\\
Apartado Postal {20 364}, 01000 M\'exico, D.F., M\'exico}

\author{Gast\'on Garc\'{\i}a-Calder\'on}
\email[Electronic mail:]{gaston@fisica.unam.mx}
\affiliation{Instituto de F\'{\i}sica,
Universidad Nacional Aut\'onoma de M\'exico\\
Apartado Postal {20 364}, 01000 M\'exico, D.F., M\'exico}

\date{\today}    

\begin{abstract}
Transient tunneling effects in triple barrier systems are investigated by considering a time-dependent solution to the Schr\"{o}dinger equation with a cutoff wave initial condition. We derive a two-level formula 
for incidence energies $E$ near the first resonance doublet of the system. Based on that expression we find that the probability density along the internal region of the potential, is governed by three oscillation  frequencies: one of them refers to the well known Bohr frequency, given in terms of the first and second resonance energies of the doublet, and the two others, represent a coupling  with the  incidence energy $E$. This allows to manipulate the above frequencies to control the tunneling transient behavior  of the probability density in the short-time regime.
\end{abstract}


\maketitle
\section{Introduction}

In this work we address the issue of time-dependent tunneling phenomena in triple barrier resonant systems, aimed to study the transient behavior of the probability density near a resonance doublet. We shall refer to these structures as  two-level {\it open} systems, in the sense that their finite-width barriers enable the system to interact with incident particles via a tunneling process. The dynamical properties of triple barrier structures have not drawn the attention they deserve. In this work we wish to emphasize that triple barriers involve novel dynamical aspects not present in double-barrier structures, where the tunneling dynamics near resonance energy is governed by a single resonance.\cite{hauge,gcr97,apl1,apl2}

The purpose of this paper is to demonstrate, based on an exact analytical approach, that the dynamics of the transient probability density is governed by three relevant frequencies that involve the resonance energies of the  doublet and the incidence energy $E$. We find that in addition to the {\it{Bohr frequency}}, ${\omega}_{21}=|{\cal E}_2-{\cal E}_1|/\hbar$, which is an intrinsic property of the system, there are two additional frequencies, 
${\omega}_{1}=|E-{\cal E}_1|/\hbar$ and ${\omega}_{2}=|E-{\cal E}_2|/\hbar$, where the resonance energies ${\cal E}_1$ and
${\cal E}_2$ are the real parts of the corresponding complex resonance energies $E_n={\cal E}_n-i \Gamma_n/2$ (n=1,2) of the problem. This should be contrasted with the well known \cite{cohen} dynamical behavior of {\it closed} two-level systems, which is governed only by the Bohr frequency, $\Omega _{12}=({\rm E}_{2}-{\rm E}_{1})/\hbar$ where ${\rm E}_{1}$ and ${\rm E}_{2}$ are the real energy eigenvalues of the system.
As shown below, the above frequencies may be manipulated to produce a significant enhancement of the short-time transient behavior of the probability density. 

The paper is organized as follows: Section II presents an overview of the formalism. In section III we discuss the transient behavior of the probability density through several numerical examples. Finally, section IV, provides some concluding remarks.

\section{The Formalism}

The model used in this work deals with an explicit solution to the time-dependent Schr\"{o}dinger equation with cutoff initial conditions, \cite{gcr97} and consists of a generalization to tunneling problems of the free quantum shutter setup that predicts {\it diffraction in time}.\cite{mm,kleber,holland} The phenomenon of diffraction of matter in time has been recently experimentally verified\cite{dalibard,hils} and has also stimulated further studies.\cite{zeilinger}  
The setup used in this work may be visualized as a quantum shutter\cite{ci} placed at $x=0$, just at the left edge of the resonant structure that extends over the interval $0\leq x\leq L $.  Upon opening the 
shutter\cite{opening} at $t=0$, the incoming initial wave, represented by a cutoff plane wave, 
\begin{equation}
\Psi (x,k;t=0)=\left\{ 
\begin{array}{cc}
e^{ikx}-e^{-ikx}, & x\leq 0 \\ 
0, & x>0,
\end{array}
\right.  
\label{1}
\end{equation}
interacts with the internal region ($0\leq x\leq L$) of the potential.
The wave solution to the time-dependent problem $\Psi (x,k;t)$ for $x >0$
and $t >0$, is given by,\cite{gcr97}
\begin{equation}
\Psi =\Phi _{k}M(y_{k})-\Phi _{-k}M(y_{-k})-\sum\limits_{n=-\infty
}^{\infty }\rho_{n}M(y_{k_{n}}).  
\label{Psiint}
\end{equation}
The quantities $\Phi_{\pm k}\equiv \Phi(x,\pm k)$ refer to the stationary
wave solution, and the factors,
\begin{equation}
\rho_{n}(x,k) \equiv 2iku_{n}(0)u_{n}(x)/(k^{2}-k_{n}^{2}),
\label{1a}
\end{equation}
are given in terms of the resonant states $\{u_{n}(x)\}$ and the complex energy eigenvalues  $E_{n}=\hbar^2k_n^2/2m$ of the problem. The complex energy eigenvalues may be written in terms of the complex wave numbers
$k_n=a_n-ib_n$, and correspond to the S-matrix poles of the problem. They are distributed in the third and fourth quadrants on the complex $k$-plane in a well known manner. The $M$-functions are defined as,\cite{gcr97} 
\begin{equation}
M(y_{s})=\frac {1}{2}w(iy_{s}),  
\label{2}
\end{equation}
where the functions $w(iy_s)$ stand for the complex error function,\cite{wiz} 
\begin{equation}
w(iy_s)={\rm e}^{y_s^2}{\rm erfc}(y_s)
\label{3}
\end{equation}
and the argument $y_{s}$, reads,
\begin{equation}
y_{s}\equiv {\rm e}^{i3\pi /4}\left( \frac {\hbar}{2m}\right)^{1/2}
 s\,t^{1/2},  
\label{4}
\end{equation}
and $s$ stands for $\pm k$ or $k_{\pm n}$. 
As shown elsewhere,\cite{gcr97} the time-dependent solution  given by 
Eq.\ (\ref{Psiint}) goes into the stationary solution $\Phi_k$ at asymptotically long times.

For triple barrier systems, the resonance spectra typically corresponds to a succession of resonance doublets, formed by the coupling of the single resonances associated with each of the two wells of the system. We shall be interested in systems where the first doublet is isolated. The approximation to Eq. (\ref{Psiint}) then reads,
\begin{eqnarray}
\Psi &\approx &\Phi_k M(y_k) -\Phi_k^{\ast} M(y_{-k})  \nonumber \\
&&-\sum\limits_{n=1}^{2} \left \{\rho _{n}M( y_{k_n}) +
\rho _{-n}M(y_{k_{-n}})\right\},
\label{twoterms}
\end{eqnarray}
where we have used $\Phi_{-k}=\Phi_k^*$. For a resonance doublet the stationary function may also be written as the sum over the first two resonance terms,\cite{overlapping} namely, 
\begin{equation}
\Phi_k (x,k)\approx \rho_{1}(x,k)+\rho_{2}(x,k),
\label{5}
\end{equation}
and consequently,
\begin{equation}
|\Phi_k|^2 \approx |\rho_{1}|^2 + |\rho_{2}|^2 + \rho_{12},
\label{5a}
\end{equation}
where $\rho_{12}=2{\rm Re}\{\rho_1 \rho_2^*\}$. Although the time-dependence of  Eq. (\ref{twoterms}) is contained in the $M$-functions, a considerable simplification of this two-level formula can be derived, in which the time dependence is explicitly given in terms of simple functions. Such a derivation is discussed in detail elsewhere\cite{yo} and we will recount it here briefly. The $M$-functions $M(y_k)$ and $M(y_{k_n})$ contained in Eq. (\ref{twoterms}), can be related to functions of the form $M(y_{-k})$ and $M(y_{k_{-n}})$ by means of the symmetry relation,\cite{gcr97} 
\begin{equation}
M(y_s)={e}^{y_s^{2}}-M(-y_s).  
\label{symmrel}
\end{equation}
Using Eq. (\ref{symmrel}) we can rewrite Eq. (\ref{twoterms}) as, 
\begin{equation}
\Psi=\sum_{n=1}^{2}\rho_{n}(x)
\left[{\rm e}^{y_k^2}-{\rm e}^{y_{k_{n}}^2}\right] +\Delta (x,t),  
\label{twoterms_b}
\end{equation}
where $\Delta (x,t)$ accounts for all the terms containing $M$ functions
of the form $M(y_{-k}) $ and $M(y_{k_{-n}})$, which behave as an inverse power of $t$, as follows from its series expansion,\cite{gcr97} $M(y_s)\sim 1/2[1/(\pi ^{1/2}y_s)-1/(\pi ^{1/2}y_s^{3})+...]$. Thus, except for extremely short or very long times compared with the lifetimes of the resonance levels of the doublet, the term $\Delta (x,t)$ gives a negligible contribution to the solution and can be neglected. By doing this, we can obtain a simple expression for the probability density, valid for the internal region and an energy $E$ close to the doublet, namely,

\begin{equation}
\left| \Psi (E,t)\right| ^{2}=\phi_{1}(E,t)+\phi_{2}(E,t)+\phi_{12}(E,t),  
\label{Nformul}
\end{equation}
where $\phi_{n}(E,t)$ and the interference terms $\phi_{mn}(E,t)$
(n=1,2) are given respectively by, 
\begin{equation}
\phi_{n}(E,t)=\left| \rho _{n}\right| ^{2}\chi _{n}(E,t),  
\label{taunE}
\end{equation}
and 
\begin{equation}
\phi_{mn}(E,t)=2
\mathop{\rm Re}
\{\rho _{m}\rho _{n}^{\ast }\xi _{mn}(E,t)\},  
\label{taumnE}
\end{equation}
where $E=\hbar ^{2}k^{2}/2m$ is the incidence energy, and the functions $\chi _{n}$ and $\xi _{mn}$ have the
following closed analytic expressions, 
\begin{equation}
\chi_{n}(E,t)=1-2\cos (\hat{\omega}_{n}t)e^{-\Gamma _{n}t/2\hbar }+e^{-\Gamma_{n}t/\hbar };  
\label{Jin}
\end{equation}
\begin{eqnarray}
\xi _{mn}(E,t) &=&[1-e^{i\hat{\omega}_{m}t-\Gamma _{m}t/2\hbar }-e^{-i\hat{\omega}_{n}t-\Gamma
_{n}t/2\hbar }  \nonumber \\
&&+e^{-i\hat{\omega}_{mn}t-(\Gamma _{m}+\Gamma _{n})t/2\hbar }].  \label{Ximn}
\end{eqnarray}
In the above expressions, $\hat{\omega}_{1}$, $\hat{\omega}_{2}$ and $\hat{\omega}_{12}$ are defined by $\hat{\omega}_{n}\equiv (E-{\cal E} _{n})/\hbar $ and $\hat{\omega}_{21}\equiv ({\cal E} _{2}-{\cal E} _{1})/\hbar$. 

The formula given by Eq. (\ref{Nformul}) is an important analytical result since it explicitly reveals novel aspects of the quantum dynamics of tunneling structures with resonance doublets. According to Eq. (\ref{Nformul}), the time-dependent probability density is the superposition of the three oscillating contributions, $\phi_{1}(E,t)$, $\phi_{2}(E,t)$ and $\phi_{21}(E,t)$, which have in general different amplitudes and frequencies. The three characteristic frequencies that govern the time evolution during the transient regime are: $\omega _{1}$, $\omega _{2}$ and $\omega _{21}$, which are respectively the absolute values of $\hat{\omega}_{1}$, $\hat{\omega}_{2}$ and $\hat{\omega}_{21}$. 
Note that at asymptotically long times, it is easily seen from Eqs.
(\ref{Jin}) and (\ref{Ximn}), respectively, that $\chi_n \rightarrow 1$
and $\xi_{mn} \rightarrow 1$, and hence the probability density for the two-level formula, given by Eq.\ (\ref{Nformul}), goes into the stationary solution given by Eq.\ (\ref{5a}).

\section{examples}
We shall be interested in analyzing the transient tunneling effects of the probability density  at the right-hand edge of the system, $x=L$,
because there appear the largest transient effects along the transmitted region. We consider as a first example a periodic triple barrier system with
parameters given as in Ref. \onlinecite{strong}, namely: barrier heights $V_{0}=0.12$ eV, barrier widths $b_{0}=3.0$ nm, well widths $w _{0}=16.0$ nm; and effective mass of the electron 
$m=0.067m_{e}$. The corresponding resonance parameters of the first doublet are: energy positions, ${\cal E} _{1}=11.512$ meV and ${\cal E} _{2}=14.387$ meV; and resonance widths, $\Gamma _{1}=0.4089$ $meV$ and $\Gamma _{2}=0.6365$ meV. \begin{figure}[tbp]
\includegraphics[width=3.4in]{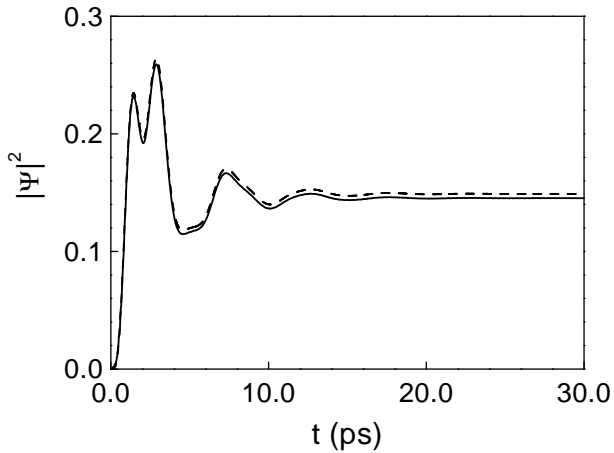}
\caption{ Time evolution of $|\Psi( L,k;t)|^{2}$ for a triple barrier system at off-incidence energy $E=\protect{\cal E}_{1}+2.0\Gamma_{1}$, using the exact solution, Eq. (\ref{Psiint}) with $N=4$ (solid line), and the two level formula, Eq. (\ref{Nformul}) (dashed line). The systems parameters are given in the text.}
\label{fig1}
\end{figure}

\begin{figure}[tbp]
\includegraphics[width=3.4in]{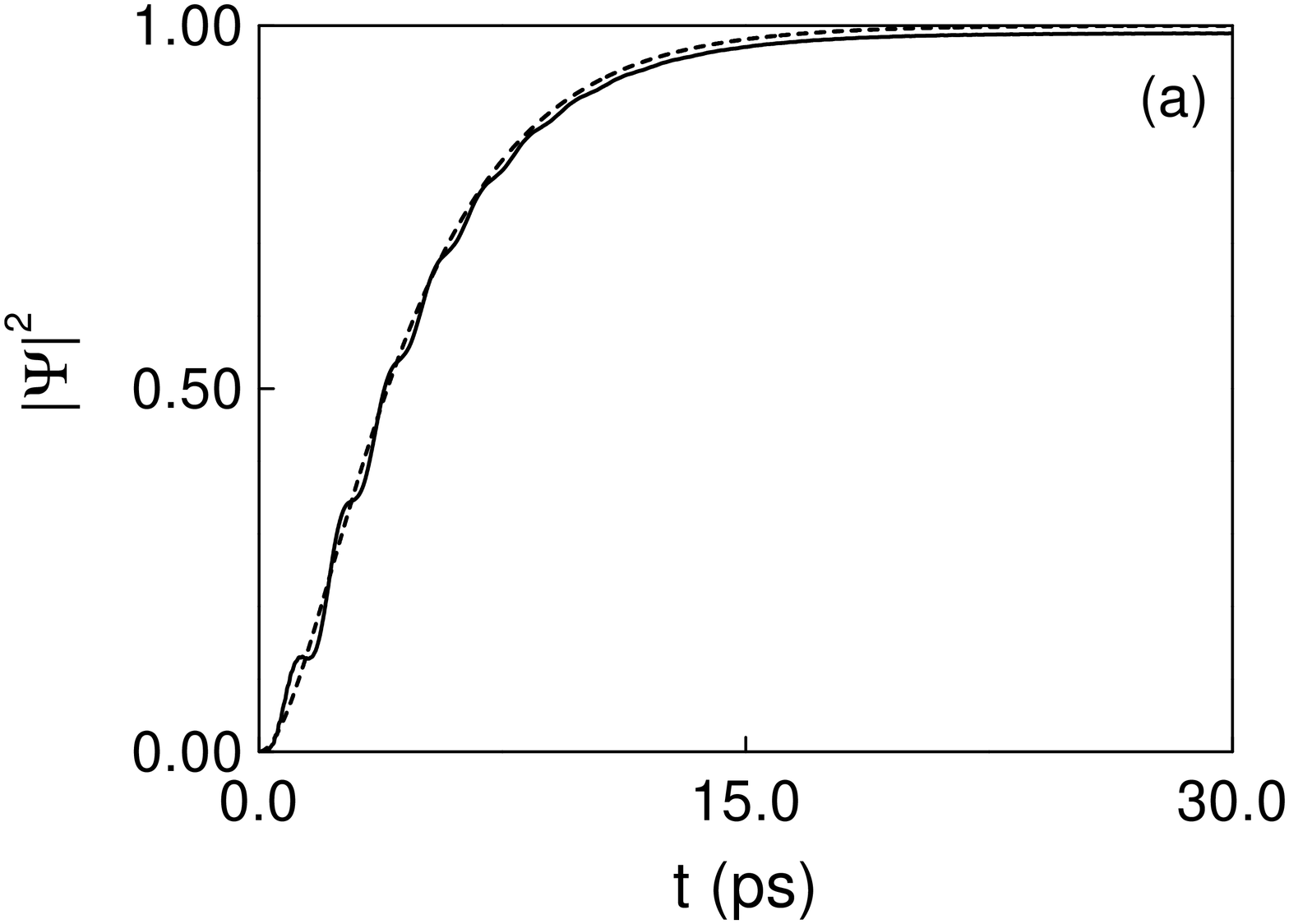}
\includegraphics[width=3.4in]{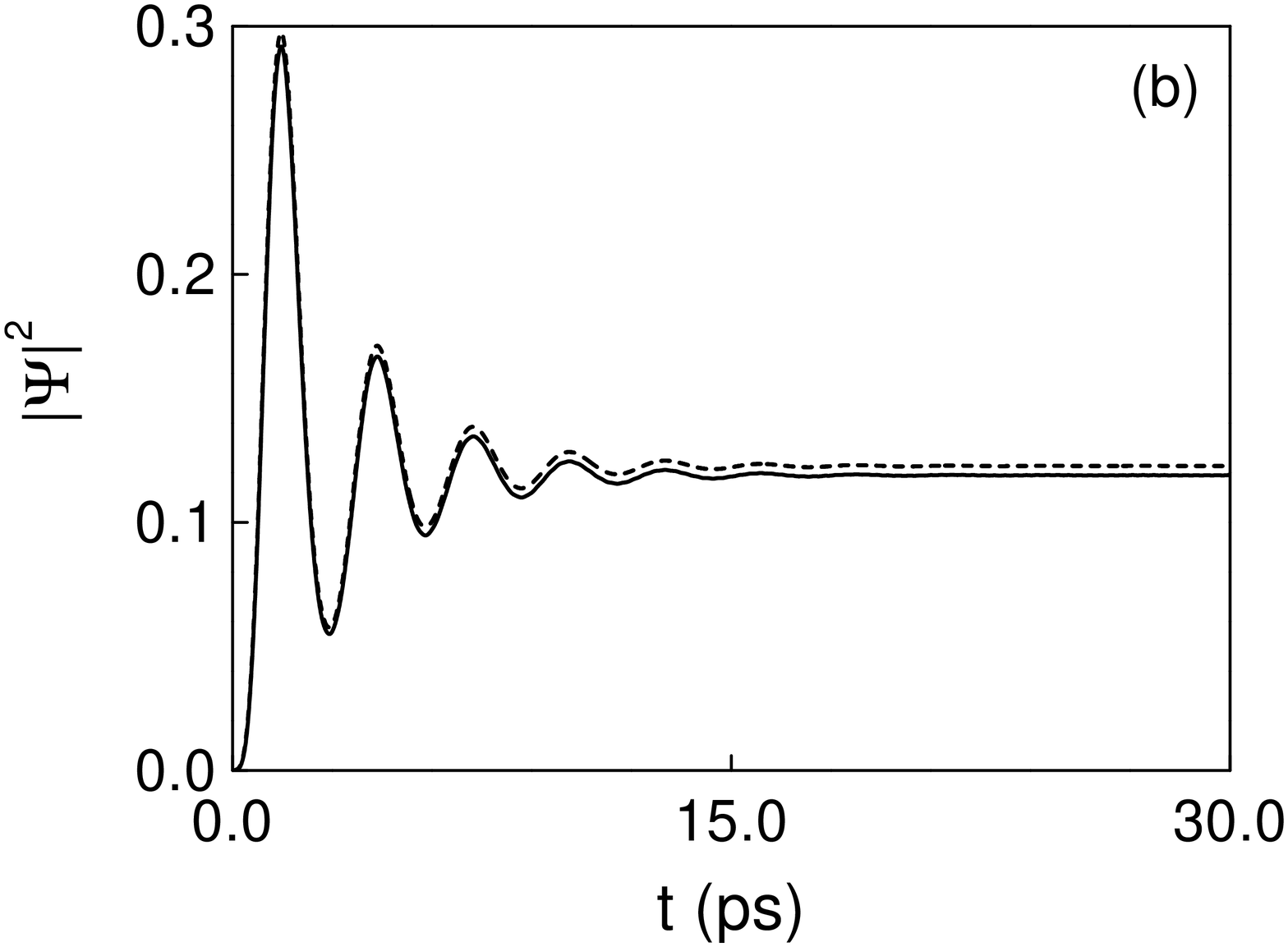}
\caption{ Time evolution of $|\Psi( L,k;t)|^{2}$ for the same triple barrier system of the previous figure, for two special situations: (a) for incidence at the first resonance, $E=\protect{\cal E}_{1}$, where $\hat{\omega}_{1}=0$ and $\hat{\omega}_{2}=-\hat{\omega}_{21}$; and (b) for incidence at $E=\bar{E}\equiv ({\cal E} _{1}+{\cal E} _{2})/2$, where $\hat{\omega}_{1}=-\hat{\omega}_{2}=1/2\hat{\omega}_{21}$. For comparison, the calculations in (a) were made by Eq. (\ref{twoterms}) (solid line) and the exponential formula given by Eq. (\ref{buildup}) (dashed line); and the calculations in (b), by Eq. (\ref{Psiint}) (solid line) and Eq. (\ref{Nformul}) (dashed line).}
\label{fig2}
\end{figure}

Let us first illustrate the reliability of our approximate formula derived above for a single doublet. In Fig. \ref{fig1} we compare the behavior of the probability density using both the formal solution, Eq. (\ref{Psiint}) (solid line), and Eq. (\ref{Nformul}) (dashed line), for an incidence energy near the first resonance, $E={\cal E}_{1}+2.0\Gamma _{1}=12.33$ meV. As can be appreciated, the two level approximation to Eq. (\ref{Psiint}) given by Eq. (\ref{Nformul}) gives an excellent description of the probability density. In this particular example, we have included in Eq. (\ref{Psiint}) the first four resonances of the systems, i.e.  $N=4$, in order to illustrate that the contribution of far away resonances is negligible. The irregular behavior of $|\Psi|^2$ observed in Fig. \ref{fig1}, arises from the interplay between $\phi_{1}$, $\phi_{2}$, and $\phi_{12}$ of Eq. (\ref{Nformul}). This situation contrasts with the regular behavior observed in double barrier structures.
As shown in a recent work,\cite{gcr97} in the case of a double barrier system the probability density grows exponentially for incidence at resonance, and exhibits regular oscillations with a single frequency if the incidence occurs near resonance.\cite{apl1} This is due to the fact that in the double barrier case, the one-level approximation stands, and hence only the term $\phi_{1}(E,t)$ is important. An interesting feature of triple barrier systems, not present in double barrier structures, is that the frequencies can be tuned by a proper choice of the incidence energy  $E$. This allows to manipulate the frequencies in such a way that the irregularities observed in Fig. \ref{fig1} dissappear. This occurs at two special situations that depend on $E$. The first situation is when the incidence energy coincides with one of the two resonances, and the second one occurs when the incidence energy coincides with the middle point between the two resonances of the doublet. In the first case, only one of the three terms of Eq. (\ref{Nformul}) dominates over the remaining two, for example if $E={\cal E}_{1}$, then $\phi_{2}(E,t)$ and $\phi_{21}(E,t)$ are negligible in comparison with $\phi_{1}(E,t)$. Since in this case $\hat{\omega}_{1}=0$, then we have that the probability density is governed by the following simple expression,
\begin{equation}
|\Psi(E={\cal E}_{1})| ^{2}\approx T({\cal E}_{1})(1-e^{-t/\tau_{1}})^{2},
\label{buildup}
\end{equation}
where $\tau_{1}=\hbar/\Gamma$ is the lifetime of the resonance $n=1$ and
$T({\cal E}_{1})$ is the peak value of the transmission coefficient, which is unity for this symmetrical system.  The results of this resonant case are depicted in Fig. \ref{fig2}(a), where we show the calculations using Equations (\ref{twoterms}) (solid line) and (\ref{buildup}) (dashed line). 
Both curves almost coincide, except by the very small oscillations of
the exact two-level formula, i.e. Eq. (\ref{twoterms}) (solid line), which are due to the effect of the second resonance of the doublet. In the second case, when the incidence energy is chosen just at the middle of the two resonances i.e., $E=\bar{E}\equiv ({\cal E} _{1}+{\cal E} _{2})/2$, we have $\bar{\omega}\equiv \hat{\omega}_{1}=-\hat{\omega}_{2}=\hat{\omega}_{21}/2$, that is, the dynamics is governed by a single frequency, $\bar{\omega}$, and the behavior of $\left| \Psi \left( L,k;t\right) \right| ^{2}$ vs $t$ is similar to a diffraction in time pattern,\cite{mm} see Fig. \ref{fig2}(b). Here the numerical value of $\bar{E}=12.949$ meV$={\cal E} _{1}+3.515\Gamma _{1}$. 

\begin{figure}[tbp]
\includegraphics[width=3.4in]{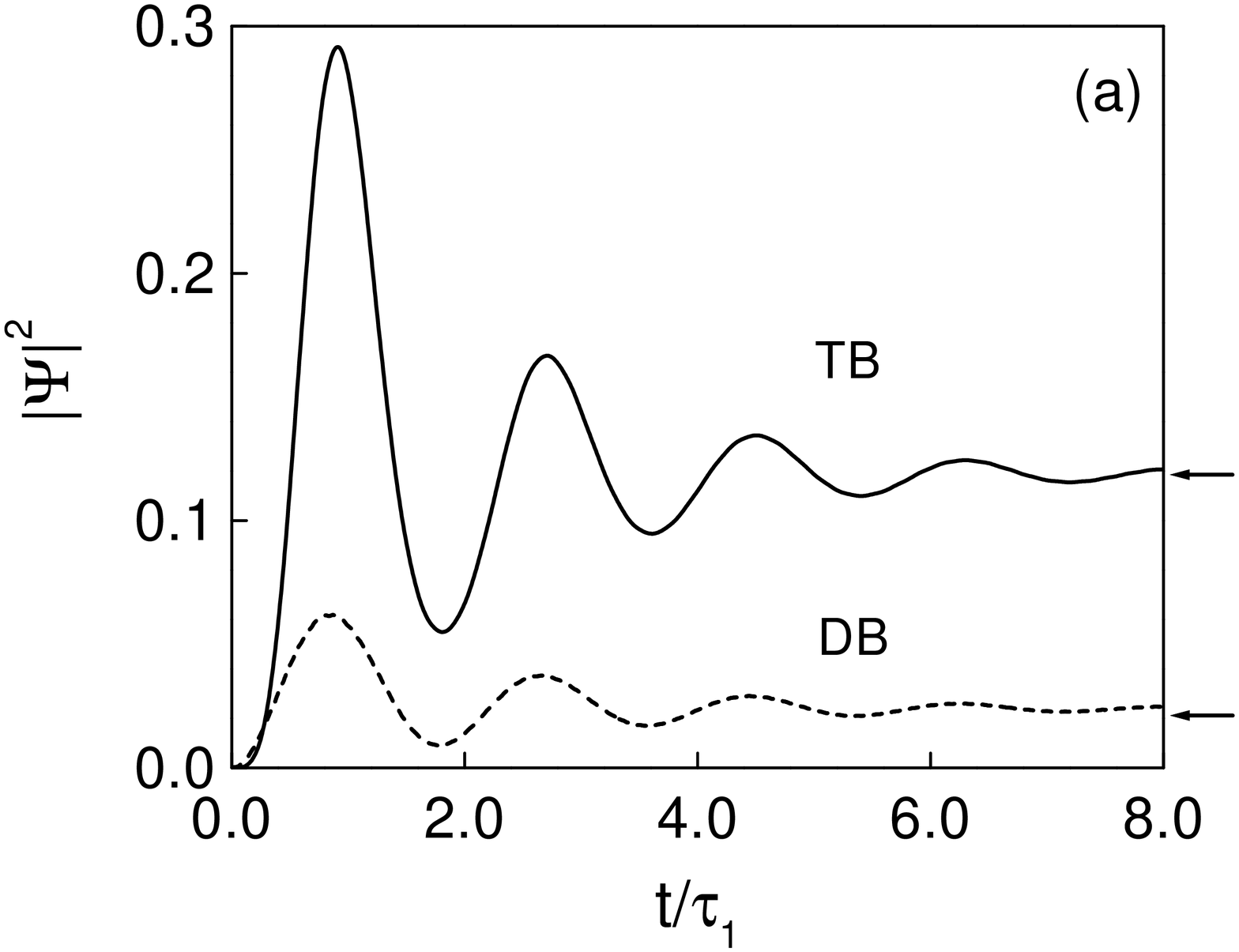}
\includegraphics[width=3.4in]{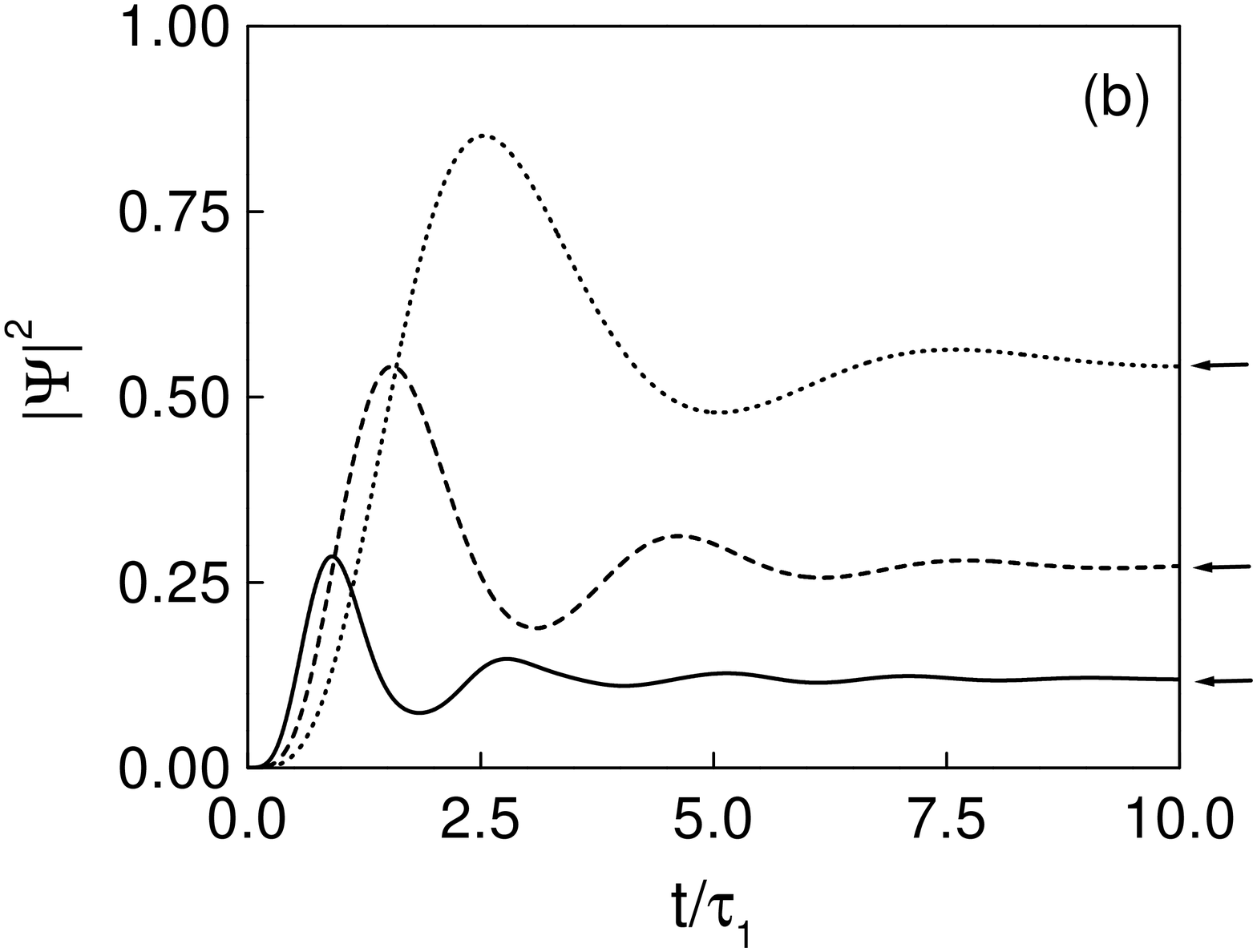}
\caption{(a) Enhancement of $|\Psi(L,k;t)|^{2}$ (solid-line) in a triple barrier systems for incidence at the center of the resonance doublet, compared with the results of a double barrier system (dashed-line). (b) Enhancement as a function of the central barrier width
$b=3.0$ nm (solid line), $b=4.0$ nm (dashed-line), and
$b=5.0$ nm (dotted-line). The arrows indicate for each case the values of the corresponding transmission coefficient. See text.}
\label{fig3}
\end{figure}
 
Compared to the double barrier case, this constructive interference effect produces an important enhancement of the transient probability density. The comparison is shown in Fig. \ref{fig3}(a) in which we used the same triple barrier structure parameters of the previous figures, and the double barrier system with potential parameters: barrier heights $V_{0}=0.23$ eV, barrier widths $b_{0}=5.0$ nm, well width $w _{0}=5.0$ nm. The first resonant state of the system has energy position, ${\cal E} _{1}=80.11$ meV; and resonance width, $\Gamma_{1}=1.033$ meV. The incidence energy was also chosen with the same deviation from resonance in units of the resonance width, that is, ${E}={\cal E} _{1}+3.515\Gamma _{1}$ whose numerical value is $83.740$ meV. Note that the scale in the time axis was normalized to the lifetime of the first resonance of each system, which for the triple barrier has the value $\tau_{1}=1.61$ ps, and for the double barrier, $\tau_{1}=6.37$ ps. Both curves tends to their correct asymptotic limits, the transmission coefficient, which for the double barrier system has the value T(E)=0.0229, and for the triple barrier system, T(E)=0.119. These values are indicated by the small arrows in Fig. \ref{fig3}(a).

As it is well known from time-independent studies in triple barrier systems,\cite{strong,yama} for incidence energies at the center of an isolated doublet, the transmission coefficient increases to unity as we increase the width $b_{2}$ of the central barrier to about twice the value of the width of the external barriers. In view that the transmission coefficient is the asymptotic value of the time dependent probability density at $x=L$, it is expected that the latter can also be enhanced in the same fashion. In Fig. \ref{fig3}(b) we illustrate how we can manipulate this extra degree of freedom to enhance the amplitude of the oscillations of the transient probability density. Here we considered $b_{2}=4.0$ nm (dashed line), and $b_{2}=5.0$ nm (dotted line); the solid line corresponds to the same triple barrier system of Fig. \ref{fig3}(a) ($b_{2}=3.0$ nm), which is also included here for comparison. Also in this figure, the values of the transmission coefficient at the energies $\bar{E}$ for each system is indicated by arrows to illustrate how the time dependent probability density tends to the correct asymptotic behavior as $t\rightarrow \infty$.

\section{concluding remarks}

In conclusion, the dynamics of the probability density for  triple barrier  resonant structures, which is a typical example of an {\it open} two-level system, has been explored. We have derived a simple analytic expression for the probability density that provides an accurate description for energies near the resonance doublet of the system. The two-level formula allows
to identify three relevant frequencies that govern the transient behavior as a function of time. The derived  formula goes into the stationary two-level solution at asymptotically long times, and thus establishes a link with the usual stationary approach. Our results suggest that the transient effects that we have discussed are of  relevance at short times and distances from the interaction region. Hopefully our results may stimulate experimental work on this subject.

\acknowledgments
The authors acknowledge financial support from Conacyt, M\'{e}xico, through
Contract No. 431100-5-32082E. One of the authors (G.G.C) acknowledges financial support
of DGAPA-UNAM under Grant No. IN101301.


\begin{thebibliography}{14}
\bibitem{ci}  The quantum shutter may be seen as a device that aids to visualize the initial condition and hence it does not form part of the potential interaction.
\bibitem{opening} The instantaneous shutter opening may be seen as `sudden approximation' to a more realistic situation.
\end{thebibliography}
\end{document}